

\def\diag{\,{\rm diag}\,}
\def\tr{\,{\rm tr}\,}
\def\rr{\right\rangle}
\def\ll{\left\langle}
\def\pmb#1{\setbox0=\hbox{$#1$}%
\kern-.025em\copy0\kern-\wd0
\kern.05em\copy0\kern-\wd0
\kern-.025em\raise.0433em\box0 }
\magnification=1200
\hoffset=-.1in
\voffset=-.2in
\vsize=7.5in
\hsize=5.6in
\tolerance 10000

\baselineskip 12pt plus 1pt minus 1pt
\pageno=0
\centerline{\bf SELF-DUAL CHERN--SIMONS SOLITONS}
\smallskip
\centerline{{\bf WITH NON-COMPACT GROUPS}\footnote{*}{This
work is supported in part by funds
provided by the U. S. Department of Energy (D.O.E.) under contract
\#DE-AC02-76ER03069, and by the Swiss National Science Foundation.}}
\vskip 24pt
\centerline{D.~Cangemi}
\vskip 12pt
\centerline{\it Center for Theoretical Physics}
\centerline{\it Laboratory for Nuclear Science}
\centerline{\it and Department of Physics}
\centerline{\it Massachusetts Institute of Technology}
\centerline{\it Cambridge, Massachusetts\ \ 02139\ \ \ U.S.A.}
\vskip 1.5in
\centerline{Submitted to: {\it Nuclear Physics B\/}}
\vfill
\centerline{ Typeset in $\TeX$ by Roger L. Gilson}
\vskip -12pt
\noindent CTP\#2092\hfill April 1992
\eject
\baselineskip 24pt plus 2pt minus 2pt
\centerline{\bf ABSTRACT}
\medskip
It is shown how to couple non-relativistic matter with a Chern--Simons gauge
field that belongs to a non-compact group.  We treat in some details the
$SL(2,{\bf R})$ and the Poincar\'e $ISO(2,1)$ groups.  For suitable
self-interactions, we are able to exhibit soliton solutions.
\vfill
\eject
\noindent{\bf INTRODUCTION}
\medskip
\nobreak
Recently,$^{1}$ there has been interest in the interaction of
non-relativistic planar matter fields with Chern--Simons gauge fields.  When
self-interactions are suitably chosen, and the gauge
group is compact, the system admits static solutions fulfilling a set of
self-dual equations.  The completeness of these static solutions has been
discussed in Ref.~[2].  A natural question addresses the
generalization to arbitrary gauge groups.  In this paper we present a general
framework that encompasses models with compact as well as non-compact and
non-semi-simple gauge groups.  In Section~I we generalize the notion of a
Killing form that we need to define the system.  In Section~II we show that
the reduction of the four-dimensional Yang--Mills self-dual equations leads to
static solutions of our problem, provided the matter fields are taken in the
adjoint representation.  In Section~III we treat as examples the semi-simple
$SL(2,{\bf R})$ group and the non-semi-simple Poincar\'e group, $ISO(2,1)$.  In
both cases, special {\it Ans\"atze\/} give explicit solutions to the static
problem.  Concluding remarks are given in Section~IV, while an Appendix
recalls some useful tools in Lie algebra theory.
\goodbreak
\bigskip
\noindent{\bf I.\quad GENERALIZATION TO NON-COMPACT LIE ALGEBRAS}
\medskip
\nobreak
Let us first recall the form taken by the Lagrangian density in the case of
compact Lie algebras ({\it i.e.\/} Lie algebra of a compact Lie group):
$$\eqalign{{\cal L} &= {\kappa\over 4}\epsilon^{\alpha\beta\gamma}\tr
A_\alpha \left( \partial_\beta A_\gamma + {1\over 3} \left[ A_\beta,
A_\gamma\right]\right) \cr
&\quad+{\rm i}  \psi^* D_t \psi - {1\over 2} \left( {\bf D}\psi\right)^* {\bf
D}
\psi - {g\over 2} \sum_a \left( \psi^* T_a\psi\right) \left( \psi^*
T_a\psi\right)\ \ .\cr}\eqno(1)$$
$\left\{ T_a\right\}$ are the generators of the algebra and $\tr T_aT_b\propto
\delta_{ab}$ is its Killing form.  As usual, $D_t = \partial_t + A_0$, ${\bf
D} = \pmb{\partial}
 - {\bf A}$ are the covariant derivatives, and the matter field
$\psi$ is an $n$-tuplet that transforms according to some finite-dimensional
representation of the group.  In the generalization of this expression, we
 preserve the two main properties of $\int d^3x\,{\cal L}$, namely its
reality and its gauge invariance.  In order to do so we replace the Killing
form and the inner product in the representing vector space by suitable
non-degenerate, Hermitian (for reality) and invariant (for gauge invariance)
bilinear forms.  Finite-dimensional
representations of non-compact groups cannot be unitary; hence we do not
expect these forms to be
positive definite.  Moreover, they do not exist in every representation (see
Appendix).

Suppose that the adjoint representation possesses such a bilinear form denoted
by $\ll T_a, T_b\rr_{\rm adj} = \Omega_{ab}$.  Suppose also
that the matter fields
belong to some representation with its own bilinear form $\ll~~,~~\rr$.  If
$\Omega^{ab}$ is the inverse matrix of $\Omega_{ab}$, the natural
generalization of Eq.~(1) is:
$$\eqalign{{\cal L} &= {\kappa\over 4}\epsilon^{\alpha\beta\gamma}\ll
A_\alpha,\partial_\beta A_\gamma + {1\over 3} \left[ A_\beta, A_\gamma\right]
\rr_{\rm adj} \cr
&\quad+{\rm i} \ll \psi,D_t\psi\rr - {1\over 2} \ll{\bf D}\psi, {\bf D}\psi\rr
-
{g\over 2} \sum\limits_{a,b} \ll \psi,T_a\psi\rr \Omega^{ab} \ll \psi, T_b\psi
\rr \ \ .\cr}\eqno(2)$$
Then most of the discussion of Ref.~[1] can be followed in this more
general case.

The equations of motion read ($\epsilon_{12}=1$):
$$\eqalignno{F^a_{xy} &= \partial_x A^a_y - \partial_y A^a_x + \left[ A_x,
A_y\right]^a = - {{\rm i}\over \kappa}
\Omega^{ab} \ll\psi, T_b\psi\rr\ \ ,&(3\hbox{a})\cr
F^a_{i0} &= {1\over 2\kappa} \epsilon_{ij} \Omega^{ab} \biggl( \ll \psi, T_b
D_j \psi\rr - \ll D_j \psi, T_b\psi\rr\biggr)\ \ ,&(3\hbox{b}) \cr
{\rm i}\partial_t\psi &= - {1\over 2} {\bf D}^2 \psi -  {\rm i}
A_0 \psi + g \ll \psi,
T_a\psi\rr \Omega^{ab} T_b\psi\ \ .&(3\hbox{c})\cr}$$
Taking Eq.~(3a) as a definition of ${\bf A}$, the last equation can also be
derived from the Hamiltonian:
$$\eqalign{H
&= {1\over 2} \int d^2{\bf r}\,\biggl( \ll {\bf D}\psi, {\bf D}\psi\rr +
g\ll \psi, T_a\psi\rr \Omega^{ab} \ll \psi, T_b\psi\rr\biggr) \cr
&= {1\over 2} \int d^2{\bf r}\, \left( \ll D_\epsilon\psi, D_\epsilon \psi\rr
+ \left( g + \epsilon {1\over \kappa}\right)\ll \psi, T_a\psi\rr \Omega^{ab}
\ll \psi, T_b\psi\rr\right)\ \ ,\cr}\eqno(4)$$
where the last equality involves the definition $D_\epsilon\equiv D_x +
{\rm i}\epsilon D_y$ ($\epsilon=\pm$) and the discarding of a boundary term.

We can list the other conserved quantities generating symmetries in the
system:$^{1}$
$$\eqalign{P^i &= \int d^2{\bf r}\, T^{0i} \cr
J &= \int d^2{\bf r}\, \epsilon_{ij} r^i T^{0j} \cr
G^i &= tP^i - \int d^2{\bf r}\,r^i\ll \psi,\psi\rr \cr
D&= tH - {1\over 2}\int d^2{\bf r}\, r^i T^{0i} \cr
K&= - t^2 H + 2tD + {1\over 2} \int d^2{\bf r}\, r^2 \ll \psi,\psi\rr \cr}
\hskip .4in
\eqalign{
\phantom{\int}&\hbox{momentum}\ \ ,\cr
\phantom{\int}&\hbox{angular momentum}\ \ ,\cr
\phantom{\int}&\hbox{Galilean boost}\ \ ,\cr
\phantom{\int}&\hbox{dilation}\ \ ,\cr
\phantom{\int}&\hbox{conformal weight}\ \ .\cr}\eqno(5)$$
In this system the momentum density $T^{0i}$ corresponds to a current:
$$T^{0i} = - {{\rm i}\over 2} \biggl( \ll \psi,D_i\psi\rr - \ll
D_i\psi,\psi\rr\biggr)\ \ .\eqno(6)$$
For static solutions, we deduce from the above that $P^i$, $D$, and
especially $H$ have to vanish.  With the special choice $g = -
\epsilon/\kappa$, the condition $H=0$ is realized if $\psi$ fulfills the
first-order differential equation:
$$D_\epsilon\psi = 0\ \ .\eqno(7)$$
It is easy to see that the solutions of Eqs.~(3a) and (7), together with:
$$A^a_0 = - \epsilon {{\rm i}\over 2\kappa} \Omega^{ab} \ll \psi,T_b \psi\rr\ \
,\eqno(8)$$
are time-independent solutions of Eq.~(3).  But, unlike for the compact case,
 the converse is not true.  Indeed,
if $\ll~~,~~\rr$ is non-positive definite (as in the non-compact case) we
cannot conclude that Eq.~(7) is the only way to achieve $H=0$ in Eq.~(4).
\goodbreak
\bigskip
\noindent{\bf II.\quad REDUCTION OF THE YANG--MILLS SELF-DUAL EQUATION}
\medskip
\nobreak
It was already pointed out that static solutions of a Chern--Simons system
are closely related to a reduction of self-dual equation expressed in four
dimensions.  This section provides a derivation of this fact for arbitrary Lie
algebras.  We consider either the ${O}(4)$ or ${O}(2,2)$ invariant
metric in order to raise and lower indices.  The self-dual Yang--Mills
equation is $\left( \epsilon^{1234} = 1 \right)$:
$$\eqalign{F^{\mu\nu} &= {1\over 2} \epsilon^{\mu\nu\alpha\beta}
F_{\alpha\beta}\ \ ,\cr
F_{\alpha\beta} &= \partial_\alpha W_\beta - \partial_\beta W_\alpha + \left[
W_\alpha, W_\beta\right]\ \ ,\cr}\eqno(9)$$
$W_\alpha$ being the gauge potentials
 with value in the Lie algebra.  The reduction
to two dimensions is achieved
by imposing translation invariance with respect to $x^3$ and
$x^4$.  Take $\kappa$ positive and write:
$$\eqalign{x&= x^1\phantom{\bigg|}\cr A_x &= W_1\phantom{\bigg|}\cr}
\hskip .4in \eqalign{y&=x^2\phantom{\bigg|}\cr A_y &= W_2\phantom{\bigg|}
\cr}\hskip .4in
\eqalign{\Psi &= \sqrt{{\kappa\over 2}} \left( W_3 - {\rm i} W_4\right) \cr
\bar{\Psi} &= - \sqrt{{\kappa\over 2}} \left( W_3 + {\rm i}W_4\right)\ \
.\cr}\eqno(10)$$
In these variables and with the definitions
$\partial_\pm = \partial_x \pm {\rm i} \partial_y$, $A_\pm = A_x \pm
{\rm i}A_y$ and $D_\epsilon$ as in Eq.~(4), Eq.~(9) reduces to:
$$\eqalignno{
\partial_- A_+ - \partial_+ A_- + \left[ A_-, A_+\right] &= {2\over\kappa}
\left[ \bar{\Psi}, \Psi\right] &(11\hbox{a})\cr
D_\epsilon \Psi &= 0\ \ ,&(11\hbox{b})\cr}$$
where in the last equation $\epsilon$ is correlated with the metric:
$\epsilon=+1$ for
${O}(4)$, and $\epsilon=-1$ for ${O}(2,2)$.

Introducing ${\cal A}_+ = A_+ + \sqrt{2/\kappa}\,\Psi$, ${\cal A}_--
\sqrt{2/\kappa}\,\bar{\Psi}$ we see that if (and only if) $\epsilon =-1$,
Eqs.~(11) are equivalent to a zero curvature condition:
$$\partial_- {\cal A}_+ - \partial_+ {\cal A}_- + \left[ {\cal A}_-, {\cal
A}_+\right] = 0\ \ .\eqno(12)$$
For compact groups, Dunne$^{2}$ has found explicitly all the solutions of
this last equation.  However, it is not clear that his construction works for
non-compact groups.
Namely, Eq.~(12) is solved in a matrix representation and the solution is a
matrix which does not necessarily belong to the algebra we are interested in.
For example, we are not able to exhibit a solution by this method in the
$ISO(2,1)$ case.

We observe that the reduced self-dual equations (11) are the same as
Eqs.~(3a) and (7) provided we take the matter fields in the adjoint
representation, {\it i.e.\/}:
$$\Psi = \sum\limits_a \psi^a T_a\ \ ,\qquad \bar{\Psi} = - \sum\limits_a
\left( \psi^a\right)^* T_a\ \ .\eqno(13)$$
Henceforth we shall work with this representation.  Moreover, in the compact
case, only the choice $\epsilon=-1$ leads to regular solutions.  In that case,
Eq.~(12) is relevant and one can follow the general discussion of Ref.~[1]
involving chiral currents and give explicit solutions.  But in the
non-compact case, different signs conspire to ensure the existence of regular
solutions only in the opposite case, $\epsilon=+1$, where Eq.~(12) is no
longer valid.  In our following illustrations we shall consider only this case.
\goodbreak
\bigskip
\noindent{\bf III.\quad SOLITON SOLUTIONS IN THE ADJOINT REPRESENTATION}
\medskip
\nobreak
We take now two examples of non-compact groups: the semi-simple $SL(2,{\bf R})$
and the non-semi-simple $ISO(2,1)$.  The matter field is in the adjoint
representation and we shall present different {\it Ans\"atze\/} to solve the
self-dual equations with $\epsilon=1$.
\goodbreak
\bigskip
\noindent{\bf III.1\quad The $\pmb{SL(2,{\bf R})}$ Case}
\medskip
\nobreak
As discussed in the Appendix, the adjoint representation carries a Killing
form $\Omega = \diag (1,-1,-1)$.  In order to get simple differential
equations from Eq.~(11), we try a solution with the gauge field ${\bf A}$ in a
maximal commutative subalgebra.  Since $SL(2,{\bf R})$ is semi-simple it does
have a decomposition of Cartan type, but due to the non-compactness, different
decompositions are not equivalent.  Indeed, we can make two choices ${\bf
A}\propto J_0$ and ${\bf A}\propto J_2$ (or $J_1$).

Let us first try the following {\it Ansatz\/}:
$$\eqalign{\Psi &= u^0 J_0 + u^+ {1\over \sqrt{2}} \left( J_1+{\rm i}
J_2\right) + u^-
{1\over \sqrt{2}} \left( J_1 - {\rm i}J_2\right)\ \ ,\cr
A_+ &= \omega J_0\ \ .\cr}\eqno(14)$$
If $u^+$ is non-zero, the self-dual equations (with $\epsilon=1$) become:
$$u^0 = 0\ \ ,\qquad \partial_+ \left( u^+u^-\right)=0\ \ ,\qquad \omega =
{\rm i}\partial_+\ln u^+\ \ ,\eqno(15\hbox{a})$$
$$\nabla^2 \ln\left|u^+\right|^2 = - {2\over \kappa} \left( \left| u^+
\right|^2 - \left| u^-\right|^2\right)\ \ .\eqno(15\hbox{b})$$
We know$^{1}$ that regular solutions are found only if $u^-=0$.  The last
equation is then the Liouville equation for the norm of $u^+$.  Its phase is
fixed (up to a gauge transformation) by requiring regularity for $\omega$.
The radially symmetric solutions are:
$$\eqalignno{u^+ &= 2\sqrt{\kappa}\, N {1\over r}\  {1\over
\left(r/r_0\right)^N
+ \left( r_0/r\right)^N} e^{{\rm i}(1-N)\theta}\ \ ,&(16\hbox{a})\cr
\omega &= - 4 {\rm i} N {1\over r}\  {\left( r/r_0\right)^N\over \left(
r/r_0\right)^N + \left( r_0/r\right)^N} e^{-{\rm i}\theta}
\ \ .&(16\hbox{b})\cr}$$
This solution carries an angular momentum $J = - \kappa 2N$ and a conformal
weight $K = - \pi\kappa r^2_0\,{\rm cosec}(\pi/N)$ (note the opposite sign
with respect to the compact $SU(2)$ case).  With the other choice for
$\epsilon$ we would have found the opposite sign in the Liouville equation
leading to no regular solution.

Consider now the other possibility for the Cartan subalgebra:
$$\eqalign{\Psi &= u^0 J_0 + u^+ {1\over \sqrt{2}} \left( J_1 + J_0\right) +
u^- {1\over \sqrt{2}} \left( J_1-J_0\right)\ \ ,\cr
A_+ &= \omega J_2\ \ .\cr}\eqno(17)$$
The self-dual equations become:
$$\eqalignno{ u^+ u^- &= C\left( x^-\right)\ \ ,&(18\hbox{a})\cr
\omega&= \partial_+ \ln u^+\ \ ,&(18\hbox{b})\cr}$$
with an arbitrary complex function $C\left( x^-\right)$.  The combination
$\phi = 2\arg u^+ + \arg C$ obeys the sine-Gordon equation in Euclidean space:
$$\nabla^2 \phi = - {2\over {\kappa}} |C| \sin\phi\ \ .\eqno(19)$$
To solve it explicitly we take $C$ constant
and we find ``multi-kink'' solutions,
regular everywhere.$^{3}$  However, they do not lead to a function
$\omega$ decreasing at infinity, unless $C=0$.  In that case the solution is:
$$\phi = \hbox{constant}\ \ ,\qquad u^+ = \left| u^+\right|
e^{{\rm i}\phi}\ \
,\qquad \omega = \partial_+ \ln \left| u^+\right|\ \ ,\eqno(20)$$
which is gauge equivalent to the trivial solution $u^+=\omega=0$ and thus
gives no new soliton solution.

Equation (16) gives regular radially symmetric solutions.  Like in the $SU(2)$
case$^2$ we expect that all of them are obtained through the {\it Ansatz\/}
(14).
\goodbreak
\bigskip
\noindent{\bf III.2\quad The $\pmb{ISO(2,1)}$ Case}
\medskip
\nobreak
The Poincar\'e group is an example of a non-compact and non-semi-simple Lie
group.  The six generators and the bilinear form $\Omega$ of the adjoint
representation are described in the Appendix.  In this non-semi-simple algebra
the Cartan decomposition has no meaning, but it is still useful to take the
gauge field in a maximal Abelian subalgebra.  Again we have two choices.  Let
us make the first {\it Ansatz\/}:
$$\eqalign{
\Psi &= u^0 J_0 + u^+ {1\over \sqrt{2}} \left( J_1 + {\rm i}J_2\right) + u^-
{1\over \sqrt{2}} \left( J_1 - {\rm i}J_2\right) \cr
&\quad + v^0 P_0 + v^+ {1\over \sqrt{2}} \left( P_1 + {\rm i}P_2\right) + v^-
{1\over\sqrt{2}} \left( P_1 - {\rm i}P_2\right) \ \ ,\cr
A_+ &= \omega J_0 + e^0 P_0\ \ .\cr}\eqno(21)$$
Always with $\epsilon=1$, the self-dual equation implies for $u^+$, $u^-$ and
$\omega$ the same equations (15) as in the previous example.  Regular
solutions were obtained only with $u^- = 0$.  With this condition the other
equations are:
$$\eqalignno{u^0 &= v^0 = 0 &(22\hbox{a})\cr
\nabla^2 \Re \left[ {v^+\over u^+}\right] &= - {2\over\kappa}\left|
u^+\right|^2 \Re \left[{v^+\over u^+}\right]\ \ ,&(22\hbox{b})\cr
e^0 &= {\rm i}\partial_+ \left( {v^+\over u^+}\right)\ \ ,&(22\hbox{c})\cr
\partial_+ \left( u^+v^-\right) &= 0\ \ .&(22\hbox{d})\cr}$$

At first sight it seems that there is not enough constraints, as Eq.~(22b)
determines the real part of $\left( v^+/u^+\right)$ but $e^0$ in Eq.~(22c)
depends also on its imaginary part.  Nevertheless, by a gauge transformation
we can always shift $e^0$ by the total derivative of a regular and real
quantity and set the imaginary part of $\left( v^+/u^+\right)$ to what we
want.  We have used the same kind of reasoning to determine the phase of
$u^+$.

We recognize Eq.~(22b) as the deformation of the Liouville equation (15b) (with
$u^-=0$).  Namely, if $\left|u^+\right|^2 = \ln \left( 1 +
\left|\phi\right|^2\right)$ is the general solution involving some analytical
function $\phi\left( x^+\right)$, we find the solutions of Eq.~(22b) by making
an arbitrary deformation $\phi\left( x^+\right) \to \phi\left( x^+\right)
\left( 1 + \epsilon \psi\left( x^+\right)\right)$:
$$\left| u^+\right|^2 \Re \left( {v^+\over u^+}\right) = \kappa\nabla^2 \left(
{|\phi|^2\over 1 + |\phi|^2}\left( \psi + \psi^*\right)\right)\ \ .\eqno(23)$$
In the ``radially symmetric'' case --- with $\phi\left( x^+\right) \propto
\left( x^+\right)^{-N}$ and $\psi\left( x^+\right)\propto \left( x^+\right)^M$
--- Eq.~(23) reads:
$$\eqalign{\Re \left( {v^+\over u^+}\right) &= {1\over N}\  {r^M\over \left(
r/r_0\right)^N + \left( r_0/r\right)^N} \cr
&\times \left[ (M-N) \left({r\over r_0}\right)^N + (M+N) \left( {r_0\over
r}\right)^N\right] \left( a_M \cos M\theta + b_M \sin M\theta\right)\ \
.\cr}\eqno(24)$$
The gauge freedom we have allows a convenient choice of its imaginary part:
$$\eqalign{ v^+ &= 2\sqrt{\kappa}\, C_{N,M} {r^{M-1}\over \left[ \left(
r/r_0\right)^N + \left( r_0/r\right)^N\right]^2} \cr
&\times \left[ (M-N) \left( {r\over r_0}\right)^N + (M+N) \left( {r_0\over
r}\right)^N\right] \, e^{{\rm i}(1-N-M)\theta}\ \ ,\cr}\eqno(25)$$
where we have used the expression (16a) for $u^+$.  Equation (22c) then gives:
$$e^0 = 4{\rm i}N\, C_{N,M} {r^{M-1}\over \left[ \left(r/r_0\right)^N + \left(
r_0/r\right)^N\right]^2}\, e^{{\rm i}(1-M)\theta}\ \ .\eqno(26)$$
In order to avoid singularities at $r=0$ and $r=\infty$ we have to restrict
the integer values of $M$ to $1-N\le M\le 1+N$.

Finally, Eq.~(22d) is trivially solved by $v^- = f\left( x^-\right)/u^+$.  As
``radially symmetric'' choice we take $f(\bar{z}) = \bar{z}^\alpha$ and
the regular solution is:
$$v^- = C_N \left[ \left( {r\over r_0}\right)^N + \left( {r_0\over
r}\right)^N \right] \left[ C_1\left( {r\over r_0}\right)^L\, e^{-{\rm i}
L\theta} +
C_2 \left( {r_0\over r}\right)^L\, e^{{\rm i}L\theta}\right]\,
e^{{\rm i}N\theta}\ \ ,\eqno(27)$$
with an integer $L\ge N$.  Equations~(16),  (25), (26) and (27) together with
$u^0 = v^0 = u^-=0$ give a soliton solution to our self-dual problem.

In fact, it is possible to consider a more general {\it Ansatz\/} with the
gauge field in a larger subalgebra than the maximal Abelian one:
$$\eqalign{\Psi &= u^+ {1\over \sqrt{2}} \left( J_1 +{\rm i}J_2\right)
+ v^0 P_0 +
v^+ {1\over \sqrt{2}} \left( P_1 + {\rm i} P_2\right)
+ v^-{1\over \sqrt{2}} \left(
P_1 - {\rm i} P_2\right)\ \ ,\cr
A_+ &= \omega J_0 + e^0 P_0 + e^+ {1\over \sqrt{2}} \left( P_1
+ {\rm i}P_2\right) +
e^- {1\over \sqrt{2}} \left( P_1 - {\rm i}P_2\right)\ \ .\cr}\eqno(28)$$
First of all we remark that since the commutators of $J_0$, $P_0$ with $P_1$,
$P_2$ only produce $P_1$, $P_2$ terms, the gauge choice
previously made for $\omega$,
$e^0$ can still be achieved.  Moreover, a gauge transformation parallel to
$P_1$, $P_2$ transforms $e^+$, $e^-$ like ($\Lambda$ is a regular complex
function):
$$\eqalign{ e^+ &\longrightarrow e^+ + \partial_+ \Lambda + {\rm i}
\omega \Lambda\ \
,\cr
e^- &\longrightarrow e^- + \partial_+ \Lambda^* - {\rm i} \omega \Lambda^* \ \
,\cr}\eqno(29)$$
while leaving $\omega$, $e^0$ unchanged.  Thus in a suitable gauge we can
also take $e^+=0$.

For $u^+$, $v^+$, $v^-$, $\omega$, $e^0$ the equations are similar to the
previous ones.  The two remaining equations for the two unknown functions
$v^0$, $e^-$ look rather simple:
$$\partial_+ \left( \partial_- v^0 - v^0 \partial_- \ln\left|
u^+\right|^2\right) = 0\ \ ,\qquad e^- = \left( u^+\right)^{-1} \partial_+
v^0\ \ .\eqno(30)$$
The first one is integrated with the help of two arbitrary functions:
$$v^0 = C_1\left( x^+\right) \left| u^+\right|^2 + \int^{x^-} dy^-\, C_2\left(
y^-\right) \left| u^+\right|^2  \left( x^+,y^-\right)\ \ .\eqno(31)$$
As an explicit example we choose $C_2=0$, a constant $C_1$ and the
``radially symmetric'' case:
$$\eqalign{
v^0 &= 4C_1\kappa N^2 {1\over r^2}\  {1\over \left[ \left( r/r_0\right)^N +
\left( r_0/r\right)^N\right]^2}\ \ ,\cr
e^- &= 2{\rm i}C_1\sqrt{\kappa}\, N{1\over r^2}\  {1\over \left[ \left(
r/r_0\right)^N + \left( r_0/r\right)^N\right]^2} \left[ (1+N) \left( {r\over
r_0}\right)^N + (1-N) \left( {r_0\over r}\right)^N\right] \, e^{{\rm i}
N\theta}\ \
.\cr}\eqno(32)$$

If we choose the gauge field in another direction in the algebra ({\it e.g.\/}
${\bf A}\propto J_2$) we would find the same trivial solution as in the
$SL(2,{\bf R})$ example.  The set of equations (16), (25), (26), (27), (32)
gives a large class of soliton solutions in the $ISO(2,1)$ case.  However, the
conserved quantities (5) give nothing interesting on these solutions.  Namely
the non-trivial ones are given here by:
$$\eqalign{ J&= - \int d^2{\bf r}\, \left| u^+\right|^2 \Re \left( {v^+\over
u^+}\right)\ \ ,\cr
G^i &= \int d^2{\bf r}\, r^i \left| u^+\right|^2 \Re \left( {v^+\over
u^+}\right)\ \ ,\cr
K&= - {1\over 2} \int d^2{\bf r}\, r^2\left| u^+\right|^2 \Re \left( {v^+\over
u^+}\right)\  \  .\cr}\eqno(33)$$
But due to the angular dependence of $\Re \left( v^+/u^+\right)$
[cf.~Eq.~(24)], $J=K=0$ and $G^i$ is non-vanishing only for $M=1$, where $e^0$
is radially symmetric.
\goodbreak
\bigskip
\noindent{\bf IV.\quad CONCLUSION}
\medskip
\nobreak
We have shown how to couple non-relativistic matter to non-compact
Chern--Simons theory.  This is not always possible since the matter field must
be in a representation that carries an invariant bilinear form.  In that case,
static equations are nicely related to the reduction of four-dimensional
Yang--Mills self-dual equations that leads to non-trivial solutions.  The
presence of these solitons can be useful to understand Euclidean gravity in
two dimensions as a reduction of a Chern--Simons system in three dimensions.
Although the matter is taken as non-relativistic, this study can also give
some insight into the question of coupling matter, in a gauge invariant form,
to $2+1$ gravity seen as a Chern--Simons theory.
\goodbreak
\bigskip
\centerline{\bf ACKNOWLEDGEMENTS}
\medskip
I thank R.~Jackiw for very helpful comments and V.~Ruuska for many discussions
about algebraic questions.
\vfill
\eject
\centerline{\bf REFERENCES}
\medskip
\item{1.}G. Dunne, R. Jackiw, S.-Y. Pi and C. Trugenberger, {\it Phys. Rev.\/}
{\bf D43} (1991) 1332; for a review see R.~Jackiw and S.-Y. Pi, in {\it
Proceedings of the Yukawa International Seminar\/}, Kyoto, Japan, 1991, to
be published.
\medskip
\item{2.}G. Dunne, ``Chern--Simons Solitons, Toda Theories and the Chiral
Model,'' MIT preprint CTP\#2079, (March, 1992), submitted to {\it
Communications in Mathematical Physics\/}.
\medskip
\item{3.}G. Leibbrandt, {\it J. Math Phys.\/} {\bf 19} (1978) 960.
\vfill
\eject
\centerline{\bf APPENDIX}
\medskip
In this Appendix we discuss the existence of non-degenerate, Hermitian and
invariant bilinear forms in a finite-dimensional representation of a
Lie algebra.
 If $u$, $v$ belong to a representing vector space and if the generators of
the Lie algebra act on them by $u\to Tu$, we are looking for a non-degenerate
bilinear form $\ll~~,~~\rr$ such that:
$$\ll u,v\rr = \ll v,u\rr^*\ \ ,\qquad \ll Tu,v\rr + \ll u,Tv\rr = 0\ \
.\eqno(\hbox{A.1})$$
In matrix notations we write $\ll u,v\rr=\left( u^m\right)^* \Omega_{mn} v^n$
with $\left(\Omega_{mn}\right)$ invertible and:
$$\Omega^\dagger = \Omega \ \  ,\qquad T^\dagger = - \Omega T\Omega^{-1}\ \
.\eqno(\hbox{A.2})$$
If the algebra is semi-simple, the adjoint representation carries such a form:
the Killing form.  But for non-semi-simple or for other representations this
is not always true.

Let us consider the following examples:
\medskip
\item{A)}Compact, semi-simple Lie algebra like $SU(n)$.  All irreducible
representations are unitary, thus in all representations $\Omega\propto I$.
\medskip
\item{B)}Non-compact, semi-simple Lie algebra. Our prototype is $SL(2,{\bf
R})$:
$$\left[ J_a, J_b\right] = \epsilon_{ab}{}^c J_c\ \ ,\eqno(\hbox{A.3})$$
with $a,b,c=0,1,2$, $\epsilon_{012} =1$, $\epsilon_{ab}{}^c = \eta^{cc'}
\epsilon_{abc'}$, $\eta^{ab} = \diag (1,-1,-1)$.  In the three-dimensional
adjoint representation we have the Killing form $\Omega = \diag(1,-1,-1)$.  In
the two-dimensional fundamental representation:
$$J_0 = {1\over 2}\left( \matrix{ 1 & \phantom{-}0\cr 0 & -1\cr}\right)\ \
,\qquad J_1 = {1\over 2} \left( \matrix{0&1\cr1&0\cr}\right)\ \ ,\qquad J_2 =
{1\over 2}\left( \matrix{\phantom{-}0 & 1 \cr-1&0\cr}\right)\ \
,\eqno(\hbox{A.4})$$
we have $\Omega = 2{\rm i}J_0$.
\medskip
\item{C)}Non-compact, non-semi-simple Lie algebra. Here we consider the
Poincar\'e algebra $ISO(2,1)$ ($a,b,c=0,1,2$):
$$\left[ J_a, J_b\right] = \epsilon_{ab}{}^c J_c\ \ , \qquad \left[ J_a,
P_b\right] = \epsilon_{ab}{}^c P_c\ \ ,\qquad \left[ P_a, P_b\right] = 0 \ \
.\eqno(\hbox{A.5})$$
In the six-dimensional adjoint representation, it turns out that there is
still a
bilinear form with the good properties (which of course is not the Killing
form):
$$\ll J_a, J_b\rr_{\rm adj} = c_1\eta_{ab}\ \ ,\quad \ll J_a, P_b\rr_{\rm adj}
= c_2 \eta_{ab}\ \ ,\qquad (c_2\not=0)\ \ ,\eqno(\hbox{A.6})$$
with $\eta_{ab}$
being the diagonal matrix $\diag(1,-1,-1)$.  But this is not true
in all representations.  For example, in the four-dimensional fundamental one
[with $\widehat{J}_a$ given by (A.4)]:
$$\eqalign{
J_a &= \left(\vcenter{\hbox{\vbox{\offinterlineskip
\def\strut{\hbox{\vrule height 8.5pt depth 4pt width 0pt}}
\halign{
\strut\hfil$#$\hfil\tabskip 0.05in&
\hfil$#$\hfil&
\hfil$#$\hfil&
\vrule#&
\hfil$#$\hfil\tabskip 0.00in\cr
 &  &  && 0 \cr
 & \hat J_a &  && 0 \cr
 &  &  && 0 \cr\noalign{\hrule}
0 & 0 & 0 && 0 \cr}}}}\right) \ \ ,\cr\noalign{\vskip 0.3cm}
P_1 &= \left(\vcenter{\hbox{\vbox{\offinterlineskip
\def\strut{\hbox{\vrule height 8.5pt depth 4pt width 0pt}}
\halign{
\strut\hfil$#$\hfil\tabskip 0.05in&
\hfil$#$\hfil&
\hfil$#$\hfil&
\vrule#&
\hfil$#$\hfil\tabskip 0.00in\cr
 &  &  && 0 \cr
 &  0 &  && 1 \cr
 &  &  && 0 \cr\noalign{\hrule}
0 & 0 & 0 && 0 \cr}}}}\right)\ \ ,\cr} \hskip .4in
\eqalign{
P_0 &= \left(\vcenter{\hbox{\vbox{\offinterlineskip
\def\strut{\hbox{\vrule height 8.5pt depth 4pt width 0pt}}
\halign{
\strut\hfil$#$\hfil\tabskip 0.05in&
\hfil$#$\hfil&
\hfil$#$\hfil&
\vrule#&
\hfil$#$\hfil\tabskip 0.00in\cr
 &  &  && 1 \cr
 &  0 &  && 0 \cr
 &  &  && 0 \cr\noalign{\hrule}
0 & 0 & 0 && 0 \cr}}}}\right)\ \ ,\cr\noalign{\vskip 0.3cm}
P_2 &= \left(\vcenter{\hbox{\vbox{\offinterlineskip
\def\strut{\hbox{\vrule height 8.5pt depth 4pt width 0pt}}
\halign{
\strut\hfil$#$\hfil\tabskip 0.05in&
\hfil$#$\hfil&
\hfil$#$\hfil&
\vrule#&
\hfil$#$\hfil\tabskip 0.00in\cr
 &  &  && 0 \cr
 &  0 &  && 0 \cr
 &  &  && 1 \cr\noalign{\hrule}
0 & 0 & 0 && 0 \cr}}}}\right)\ \ ,\cr}\eqno(\hbox{A.7})$$
we cannot find an invertible
 matrix $\Omega$ with the properties (A.2).  On the other
hand, there is another four-dimensional representation given in terms of
$4\times 4$ gamma matrices $\Gamma^A$ ($A=0,1,2,3)$ of the four-dimensional
Minkowskian space ($\Gamma_A = \eta_{AB}\Gamma^B$, $\Gamma^5 = {\rm i}\Gamma^0
\Gamma^1\Gamma^2\Gamma^3$):
$$\eqalign{J_a &= - {1\over 4} \epsilon_{abc}\Gamma^b \Gamma^c\ \ ,\cr
P_a &= {\rm i}\beta \Gamma_a \left( 1 + \Gamma^5\right)\ \ ,\qquad
(\beta\in{\bf
R})\ \ ,\cr}\eqno(\hbox{A.8})$$
which carries the bilinear form $\Omega = \Gamma_0$.
\par
\vfill
\end